\documentclass{aa}

\usepackage{graphicx}
\usepackage{txfonts}
\usepackage{wrapfig}
\usepackage{lscape}
\usepackage{rotating}
\usepackage{array}
\usepackage{colortbl}
\usepackage{xcolor}
\usepackage{gensymb}
\usepackage{ulem}
\usepackage[rightcaption]{sidecap}
\sidecaptionvpos{figure}{c}
\usepackage[export]{adjustbox}
\usepackage[draft,pdftex,pdfpagemode={UseOutlines},bookmarks,bookmarksopen,colorlinks, linkcolor={magenta},citecolor={blue},urlcolor={blue}]{hyperref}
\DeclareGraphicsExtensions{.png}

\newcommand{\ks}{\ifmmode K{\rm s} \else $K$s\fi\xspace}
\newcommand{\aks}{\ifmmode A_{K{\rm s}} \else $A_{K\mathrm{s}}$\fi\xspace}
\newcommand{\jk}{\ifmmode J - K_{\rm s} \else $J - K_{\rm s}$\fi\xspace}

\defcitealias{Bailer-Jones2018AJ....156...58B}{BJ18}

\begin{document}

     \title{VVV survey near-infrared colour catalogue of known variable stars\thanks{Based on observations taken  within the ESO Public Survey, Programme IDs 179.B-2002.}\thanks{Data used in this work is fully and only available in electronic form at the CDS via anonymous ftp to cdsarc.u-strasbg.fr (130.79.128.5) or via http://cdsweb.u-strasbg.fr/cgi-bin/qcat?J/A+A/}}

%\subtitle{}

\author{F.~R.~Herpich\inst{1}, C. E. Ferreira Lopes\inst{2}, R.~K.~Saito\inst{3}, D.~Minniti\inst{4,5}, A.~Ederoclite\inst{1}, T.~S.~Ferreira\inst{3}, M. Catelan\inst{6,7}
}

\institute{Universidade de S\~ao Paulo, IAG, Rua do Mat\~ao 1226, Cidade
Universit\'aria, S\~ao Paulo 05508-900, Brazil
%              \email{wuchterl@amok.ast.univie.ac.at}
        \and
        National Institute For Space Research (INPE/MCTI), Av. dos Astronautas, 1758 - São Jos\'e dos Campos - SP, 12227-010, Brazil
         \and
         Departamento de  F\'{i}sica, Universidade  Federal de Santa Catarina, Trindade 88040-900, Florian\'opolis, SC, Brazil
%             \email{c.ptolemy@hipparch.uheaven.space}
%             \thanks{The university of heaven temporarily does not
%                     accept e-mails}
         \and 
         Departamento  de  Fisica,  Facultad  de  Ciencias  Exactas,
  Universidad  Andres Bello,  Av.  Fernandez  Concha 700,  Las Condes,
  Santiago, Chile
         \and  
        Vatican Observatory, V00120 Vatican City State, Italy
        \and
        Instituto de Astrofísica, Facultad de F\'isica, Pontificia Universidad Cat\'olica de Chile, Av. Vicu\~na Mackenna 4860, 7820436 Macul, Santiago, Chile
        \and
        Millennium Institute of Astrophysics, Santiago, Chile
             }

   %\titlerunning{A VVV survey catalogue}
   \authorrunning{Herpich et al.}
   
   \date{Received month day, year; accepted month day, year}

% \abstract{}{}{}{}{} 
% 5 {} token are mandatory
 
   \abstract
   {The Vista Variables in the Via Lactea (VVV) near-infrared variability survey explores some of the  most complex regions of the Milky
     Way  bulge  and  disk  in  terms  of  high  extinction  and  high
     crowding.}
   {We  add a  new  wavelength dimension  to  the optical  information
     available at the American Association of Variable Star Observers International Variable Star Index (VSX-AAVSO) catalogue to test the  VVV survey  near-infrared  photometry to better characterise these objects.}
   {We  cross-matched the VVV and the VSX-AAVSO
     catalogues along with Gaia Data Release 2 photometry and parallax.}
   {We   present  a catalogue that   includes  accurate   individual
     coordinates,   near-infrared    magnitudes   ($ZYJH\ks$),   extinctions
     $A_{Ks}$, and distances based on Gaia parallaxes.   We also show
     the  near-infrared CMDs  and  spatial distributions  for the  different
     VSX types of variable stars,  including important distance indicators,
     such as RR Lyrae, Cepheids, and Miras. By analysing the photometric flags in our catalogue, we found that about 20\%\ of the stars with measured and verified variability are flagged as non-stellar sources, even when they are outside of the saturation and/or noise regimes. Additionally, we pair-matched our sample with the VIVA catalogue and found that more than half of our sources are missing from the VVV variability list, mostly due to observations with low signal-to-noise ratio or photometric problems with a low percentage due to failures in the selection process.}
   {Our results  suggest that the current  knowledge of the variability
     in the Galaxy  is biased to nearby stars with low  extinction. The
     present catalogue also  provides the groundwork for  characterising the
     results  of future  large variability  surveys such as  the Vera C. Rubin Observatory Legacy Survey of Space and Time  in  the
     highly crowded and reddened regions of the Galactic plane, as well as follow-up campaigns for characterising specific types of variables. The analysis of the incorrectly flagged stars can be used to improve the photometric classification of the VVV data, allowing us to expand the amount of data considered useful for science purposes. In addition, we provide a list of stars that are missed by the VIVA procedures for which the observations are good and which were missed due to some failure in the VIVA selection process.}
     \keywords{Galaxy: disk -- Galaxy: bulge -- Galaxy: stellar content -- Stars: variables: general}

   \maketitle
%
%-------------------------------------------------------------------

\section{Introduction}

The variable stars in the  Milky  Way are almost
countless. Since the discovery of the first of these objects, the number of variable stars has increased consistently. \citet{Pigott1786RSPT.76.189P} provided one of the first
catalogues of stellar variability, containing a total of  12 objects known
as being variable at that time. Many other  catalogues have  been released
since then, especially within the past two decades, when
large digital  survey telescopes became operational,  scanning the  sky night  after
night,  such   as the Infrared Astronomical Satellite \citep[IRAS,][]{Neugebauer1984ApJ.278L.1N}, the Two Micron All-Sky Survey
\citep[2MASS,][]{Kleinmann1994ApSS.217.11K}, the Sloan Digital Sky Survey \citep[SDSS,][]{York2000AJ.120.1579Y},
the Catalina Sky Surveys \citep{2014ApJS..213....9D,2017MNRAS.469.3688D}, the Panoramic Survey Telescope and Rapid Response System
\citep[Pan-STARRS,][]{Kaiser2002SPIE.4836.154K}, the UKIRT Infrared Deep Sky Survey
\citep[UKIDSS,][]{Lawrence2007MNRAS.379.1599L}, the AKARI Far-infrared All-Sky Survey 
\citep[AKARI,][]{Ishihara2010AA.514A.1I},  the Wide-field Infrared Survey Explorer
\citep[WISE,][]{Wright2010AJ.140.1868W,2018ApJS..237...28C,Chen2018ApJS..237...28C},
and         the           Gaia         spectroscopic         survey
\citep[Gaia,][]{Gilmore2012Msngr.147.25G,GaiaMission2016A&A...595A...1G}, as well the ongoing All-Sky Automated Survey for Supernovae \citep[ASAS-SN;][]{Jayasinghe2018MNRAS.477.3145J}. 
In addition, the large microlensing surveys, such as the Massive Astrophysical Compact Halo Objects \citep[MACHO;][]{Alcock1993Natur.365..621A}, the Optical Gravitational Lensing Experiment \citep[OGLE;][]{Udalski1993AcA....43..289U},  the Experience pour la Recherche d’Objets Sombres \citep[EROS;][]{Aubourg1993Natur.365..623A}, the Microlensing Observations in Astrophysics \citep[MOA;][]{Bond2001MNRAS.327..868B}, the Disk Unseen Objects \citep[DUO;][]{Alard1995Msngr..80...31A}, and the Korea Microlensing Telescope Network \citep[KMTNet;][]{Kim2016JKAS...49...37K}, have discovered hundreds of thousands of variable stars in the past decades.

These projects  provide a new view of the  variability in the
Galaxy, and this subject is expected to become yet more interesting.  The
future of this field of study in the Milky Way will be revolutionised
by the  Vera C. Rubin Observatory Legacy Survey of Space and Time (LSST\footnote{http://adsabs.harvard.edu/abs/2009arXiv0912.0201L},
\citealt{Ivezic2008SerAJ.176....1I}), which is expected to discover millions of variables.  Another important aspect to consider is the contrast between near-infrared and optical variability. The searches performed  at different wavelengths  result in  different relative  numbers, and this must be  considered when the number of variable stars that will be present in the LSST catalogues is predicted \citep[e.g.][]{Pietrukowicz2012AA...537A.116P}.

However, most of the current  projects searching for variables operate
with optical  telescopes, thus avoiding  the innermost Milk Way plane, where high extinction  and crowding limit the  depth at optical wavelengths. Therefore an infrared survey is more suitable for the mission  of observing deeper  into the  Galaxy plane, as  does the
VISTA         Variables       in        The       V\'ia        L\'actea
(VVV\footnote{\url{https://vvvsurvey.org/}}),  which  is  an  European
Southern Observatory  (ESO\footnote{\url{http://www.eso.org/}}) public
survey  that  mapped  the  bulge  ($-10.0^\circ  \lesssim  l  \lesssim
+10.5^\circ$ and $-10.3^\circ \lesssim b \lesssim +5.1^\circ$) and the
inner  southern part  of the  disk ($294.7^\circ  \lesssim l  \lesssim
+350.0^\circ$ and  $-2.25^\circ \lesssim  b \lesssim  +2.25^\circ$) of
our Galaxy using five near-infrared bands \citep[$Z$, $Y$, $J$,
  $H$ and $K_{\rm  s}$;][]{Minniti2010NewA.15.433M} plus a variability
campaign in $K_{\rm s}$ band with typically 100 epochs per field in
the period  2010-2017. The VVV Survey  was completed in 2017  when its
extension,  the   VVV  eXtended   Survey  \citep{2018ASSP...51...63M},
started  observing  to increase  the observed
area of VVV  from 562 to 1700  sq. deg., filling the  gaps between the
VVV        and        the        VISTA        Hemisphere        Survey
\citep[VHS][]{2013Msngr.154...35M}.

Near-infrared surveys  such as the  VVV are  very efficient at  low Galactic
latitudes  close  to the  plane,  where  existing optical  variability
surveys  are usually  blinded by  the absorption  in the  interstellar
medium. For instance, VVV data were used to find RR Lyrae stars within
100 arcmin from the Galactic Centre \citep{2018ApJ...863...79C}. In the same way that Baade's  Window was important for optical studies
of the Galactic bulge, the near-infrared  surveys also profit from the study
of the  windows that have recently been found in the  Milky Way plane
\citep[e.g.  Dante's Window;][]{Minniti2018AA...616A..26M}.

Focusing on the subject of  variables stars, the International Variable
Star Index  (VSX), provided  by the  American Association  of Variable
Star   Observers   (AAVSO)\footnote{\url{https://www.aavso.org/vsx/}},
compiles  a large  number  of  known variable objects  into a  single
database containing 1\,432\,563 objects  as of
February 1, 2020\footnote{We downloaded the present  VSX catalogue from the  AAVSO database in February 2020. As  this catalogue grows constantly, it is expected to be even larger at the time of publishing.}.  This  work compiles information  from  various catalogues  such as those  mentioned before, together with a brief analysis of missed sources in one of the more complete variability catalogues for the galactic bulge existing to date, the VISTA Variables in the Vía Láctea infrared variability catalogue \citep[VIVA,][]{FerreiraLopes2020MNRAS.496.1730F}.

Here we present a catalogue with near-infrared colours for the VSX sources based on VVV data.  This work provides useful information such as the colours in the  $ZYJHK_{\rm s}$ bands, the extinction in the near-infrared  (\aks), and the distances based on the parallaxes of  Gaia Data Release 2 \citep[Gaia DR2;][]{Gaia2018AA...616A...1G,GaiaDR22019A&A...623A.110G}.  Near-infrared colour-magnitude diagrams  (CMDs) and the surface density distribution for the different types of variable stars include important distance indicators such as RR Lyrae, Cepheids and Miras.  Section~\ref{sec:data} presents the data we used and some statistics of our sample, in Section~\ref{sec:varGaia} we classify the variable stars in our catalogue in context with distance and period ($P$), in Section~\ref{sec:viva} we discuss the sources that were missed in the construction of the VIVA catalogue, and in Section~\ref{sec:conclude} we summarise our results.

\section{Data}\label{sec:data}

Many catalogues of variable stars are available electronically, such
as    the    General    Catalogue    of    Variable    Stars    (GCVS,
\citealt{Samus2017ARep.61.80S}), the  All-Sky Automated  Survey (ASAS,
\citealt{Pojmanski2005AcA.55.275P}), and  the  Catalog  and  Atlas  of
Cataclysmic   Variables   \citep{Downes2001PASP.113.764D}. As part of this sample of catalogues, objects  are discovered by important variability surveys of the inner Milky Way such as OGLE \citep[e.g.][]{Udalski2015AcA.65.1U}, MOA \citep{Alcock1997ApJ.491.436A} , and  even   VVV.   The  VSX  catalogue contains       names,      positions,       period,      the       VSX
type\footnote{\url{https://www.aavso.org/vsx/index.php?view=about.vartypes}},
and astronomical information, such as the constellation they belong to, and the passband that was used to measure the variability, which is mostly observed in the optical.

\subsection{Cross-matching the VVV data with the VSX catalogue}\label{sec:match}

While the VSX catalogue contains data for the whole sky, the VVV survey covered a total  of 562 sq.  deg., which is about 1.4\,$\%$ of  the celestial sphere. Although it covers only a  small fraction of
the sky,  the VVV  concentrated its observations  on the  most crowded
regions of  the southern  sky, which  are the Milk  Way  bulge and
southern plane. The projected cone  of view covers $\sim$30\,$\%$
of the Milk Way stars, thus providing one of the most complete catalogues of
the   inner    Milk Way:      almost   a   billion    sources   have been detected
\citep{Alonso-Garcia2018A&A...619A...4A}.

We matched the VSX catalogue, containing 1\,432\,563 sources, with the VVV catalogues for the 348 individual tiles covering the bulge and disk portions, which contain 428\,260\,599 sources. This resulted in 701\,256 objects that match to within 1 arcsec. We refer to them as the V$^4$SX sample. Table \ref{tab:matchedtable}  shows the first  few stars  of the
resulting  match (the  full table  is available  in the  complementary
data)\footnote{In Table \ref{tab:matchedtable} some VSX and Gaia (see Section \ref{sec:matchgaia})  columns were suppressed to  better fit the page.}.  The full description of all columns appearing in the table is given in Section~\ref{sec:coldesc}. VVV  data include the coordinates (equatorial  and Galactic)  and the standard  single-epoch aperture  photometry for the VVV Data Release 4 (VVV-DR4) provided  by the Cambridge Astronomical Survey Unit (CASU\footnote{\url{http://casu.ast.cam.ac.uk/}}) in  the  five  VISTA  filters
($ZYJH\ks$), with  photometric  flags  indicating  the  likely
morphological   type   based  on   the   aperture   curve  of   growth
\citep{Saito2012AA.537A.107S}. The \ks-band values are mean magnitudes calculated from the multi-epoch observations (typically 50-100), while the $ZYJH$-band magnitudes are averages of a few (typically 2-4) observations taken at random phases for the periodic variables. The  CASU  reduction process produces a  flag to identify the quality of  the source, where
-1 means best quality photometry of stellar objects, -2 means borderline stellar
objects, 0 means noise, +1 indicates non-stellar objects, -7 indicates sources containing bad
pixels,  and -9 means saturated sources  \citep{Saito2012AA.537A.107S}. Using these flags to select unsaturated, noisy, and extended sources in the $J\ks$ bands, we  collected a sample of 281\,536 variables stars  that can be  reliably studied and whose unsaturated
light curves are available  through the VISTA Science Archive (VSA)  (we call these  CVVS,  for ``constrained VISTA  variable sources'').

While this paper focuses on the near-infrared colours, VVV  $\ks$-bands light curves are publicly available and can be retrieved through the VSA by querying the  VVV  DR4 synoptic source table\footnote{\url{http://horus.roe.ac.uk/vsa/index.html}}.  A search for new  variable stars in the VVV  light curves was performed using different techniques \citep[e.g.][]{2016AA...586A..36F,2017AA...604A.121F,FerreiraLopes2018MNRAS.481.3083F} and resulted in the discovery of a large number of previously unknown variable stars. These discoveries are beyond the scope of this paper, and the resulting catalogue is presented in \citet{FerreiraLopes2020MNRAS.496.1730F}.

\begin{table*}[!ht]
\centering
\caption{V$^4$SX catalogue. In  this sample table we  suppressed a few
  columns from  the original AAVSO  catalogue (AUID, Period,  Type, and
  Mag.),  VVV (errors  and flags),  and from  Gaia (parallax).   RA/DEC
  coordinates are J2000  from VVV. Blank entries indicate unavailable data.}
\vspace{2mm}
\label{tab:matchedtable}
\begin{tabular}{cccccccccccccc}
\hline\hline
VSX  & RA    & DEC   & L      & B      & $Z$     & $Y$     & $J$     & $H$     & $K_{\rm s}$    & $A_{Ks}$ & Dist\_BJ \\
Name & (deg) & (deg) & (deg)  & (deg)  & (mag) & (mag) & (mag) & (mag) & (mag) & (mag) &  (kpc)  \\
\hline%\hline
ASAS J123010-6323.5  &  187.54 & -63.39 & -59.45 & -0.62 & 10.93 & 10.74 & 10.58 & 10.47 & 10.34 & 0.65 &  ... \\
GDS\_J1229509-635129 &  187.46 & -63.86 & -59.44 & -1.09 & 12.72 & 11.84 & 10.40 & 10.25 &  9.01 & 0.78 &  ... \\
NSV 5667             &  187.73 & -62.77 & -59.41 &  0.01 & 13.35 & 13.15 & 12.82 & 12.96 & 12.54 & 1.08 & 0.790\\
GDS\_J1230188-633823 &  187.58 & -63.64 & -59.41 & -0.86 & 11.14 & 10.69 & 10.12 &  9.72 &  9.37 & 0.67 & 1.623\\
GDS\_J1230367-632806 &  187.65 & -63.47 & -59.39 & -0.69 & 13.50 & 13.21 & 12.83 & 12.38 & 12.13 & 0.65 & 1.696\\
GDS\_J1231003-631727 &  187.75 & -63.29 & -59.36 & -0.51 & 12.93 & 10.79 &  9.07 &  8.59 &  7.62 & 0.72 & 5.226\\
GDS\_J1230537-633727 &  187.72 & -63.62 & -59.35 & -0.84 & 10.83 & 9.16  &  8.38 &  9.29 &  8.39 & 0.67 & 1.849\\
NSV 5671             &  187.77 & -63.38 & -59.35 & -0.60 & 12.56 & 12.33 & 12.05 & 11.80 & 11.62 & 0.65 & 0.607\\
NSV 5675             &  187.88 & -62.93 & -59.33 & -0.15 & 12.65 & 12.34 & 12.24 & 12.72 & 12.23 & 0.91 & 2.128\\
GDS\_J1231035-633759 &  187.76 & -63.63 & -59.33 & -0.85 & 11.85 & 10.24 &  9.04 &  8.70 &  ...  & 0.88 & 3.162\\
GDS\_J1231255-633150 &  187.86 & -63.53 & -59.30 & -0.75 & 12.62 & 12.33 & 11.88 & 11.57 & 11.36 & 0.72 & 0.846\\
GDS\_J1231445-631137 &  187.94 & -63.19 & -59.29 & -0.41 & 12.07 & 11.88 & 11.49 & 11.37 & 11.07 & 0.84 & 0.792\\
GDS\_J1232008-624848 &  188.00 & -62.81 & -59.28 & -0.03 & 12.34 & 12.20 &  ...  &  ...  &  ...  & 0.96 &  ... \\
NSV 5684             &  187.97 & -63.21 & -59.27 & -0.42 & 13.40 & 13.17 & 12.85 & 12.56 & 12.38 & 0.84 & 0.839\\
GDS\_J1232131-624557 &  188.05 & -62.77 & -59.26 &  0.02 & 12.29 & 12.16 & 12.80 &  ...  &  ...  & 0.96 & 1.932\\
NSV 5689             &  188.03 & -63.07 & -59.25 & -0.28 & 13.27 & 13.06 & 12.75 & 12.74 & 12.45 & 0.84 & 1.316\\
ASAS J123144-6324.3  &  187.93 & -63.41 & -59.27 & -0.62 & 10.09 &  9.89 &  ...  &  ...  &  ...  & 0.72 & 1.215\\
GDS\_J1231514-633314 &  187.96 & -63.55 & -59.25 & -0.77 & 11.92 & 10.31 &  8.98 &  ...  &  7.58 & 0.88 & 4.274\\
NSV 5683             &  187.96 & -63.83 & -59.23 & -1.04 & 12.75 & 12.31 & 11.89 & 11.60 & 11.30 & 1.03 & 2.327\\
GDS\_J1232312-625756 &  188.13 & -62.97 & -59.22 & -0.17 & 12.36 & 11.79 & 12.03 & 13.92 & 11.76 & 0.91 & 3.955\\
GDS\_J1232281-630555 &  188.12 & -63.10 & -59.21 & -0.31 & 13.56 & 11.66 & 10.14 & 12.43 &  8.57 & 0.84 & 2.840\\
NSV 5692             &  188.14 & -63.23 & -59.19 & -0.44 & 13.33 & 13.15 & 12.89 & 12.67 & 12.60 & 0.79 & 1.319\\
NSV 5696             &  188.20 & -63.10 & -59.18 & -0.30 & 13.62 & 13.47 & 13.17 & 13.10 & 12.88 & 0.84 & 1.431\\
GDS\_J1232513-630146 &  188.21 & -63.03 & -59.17 & -0.23 & 13.94 & 13.70 & 13.59 & 13.46 & 13.10 & 0.91 & 0.821\\
NSV 5710             &  188.33 & -63.02 & -59.12 & -0.22 & 12.72 & 12.56 & 12.27 & 12.81 & 12.30 & 0.91 & 0.932\\
NSV 5701             &  188.24 & -63.81 & -59.11 & -1.01 & 17.79 & 16.16 & 14.28 & 12.35 & 11.55 & 0.82 & 3.727\\
ASAS J123326-6256.2  &  188.36 & -62.94 & -59.11 & -0.14 & 12.90 & 12.40 & 13.77 &  ...  &  ...  & 0.91 & 1.426\\
NSV 5699             &  188.24 & -63.85 & -59.10 & -1.05 & 11.08 & 10.84 & 10.45 & 10.07 &  9.87 & 0.82 &  ... \\
NSV 5713             &  188.37 & -63.26 & -59.08 & -0.46 & 13.51 & 13.29 & 12.99 & 13.20 & 12.96 & 0.71 & 0.971\\
VW Cru               &  188.33 & -63.51 & -59.09 & -0.71 & 13.32 & 13.00 & 13.74 &  ...  &  ...  & 0.72 & 1.235\\
CM Cru               &  188.48 & -62.83 & -59.07 & -0.03 & 14.36 & 13.49 & 12.52 & 12.66 &  ...  & 0.93 & 3.656\\
GDS\_J1233486-630008 &  188.45 & -63.00 & -59.07 & -0.20 & 12.34 &  ...  & 12.74 &  ...  &  ...  & 0.91 & 2.442\\
GDS\_J1233485-630858 &  188.45 & -63.15 & -59.06 & -0.35 & 12.27 & 12.08 &  ...  &  ...  &  ...  & 0.72 & 9.390\\
GDS\_J1233514-631706 &  188.46 & -63.29 & -59.04 & -0.48 & 12.29 & 11.93 & 12.05 &  ...  &  ...  & 0.71 & 6.043\\
NSV 5716             &  188.43 & -63.82 & -59.02 & -1.01 & 12.00 & 11.57 & 11.20 & 10.99 & 10.79 & 0.82 & 1.587\\
GDS\_J1234027-632932 &  188.51 & -63.49 & -59.01 & -0.69 & 12.03 & 11.90 & 12.33 &  ...  &  ...  & 0.72 & 1.715\\
GDS\_J1233526-635121 &  188.47 & -63.86 & -59.00 & -1.05 & 11.78 & 11.26 & 10.71 & 10.19 & 11.41 & 0.68 & 0.544\\
NSV 5719             &  188.46 & -63.93 & -59.00 & -1.13 & 12.96 & 12.64 & 12.37 & 12.08 & 11.83 & 0.51 & 0.901\\
GDS\_J1234165-631955 &  188.57 & -63.33 & -58.99 & -0.52 & 12.35 & 11.85 & 11.91 &  ...  &  ...  & 0.69 & 6.543\\
GDS\_J1234253-630811 &  188.61 & -63.14 & -58.99 & -0.33 & 12.34 & 11.92 & 12.13 &  ...  &  ...  & 0.83 & 5.700\\
\hline
\end{tabular}
\end{table*}

\subsection{Extinction in the near-infrared from the VVV maps}\label{sec:aks}

Complementing the near-infrared colours from the VVV, we also present the total extinction in the \ks\  band (\aks) in our catalogue, integrated along the entire line of sight for each source and provided by the  VVV extinction maps.   In  all cases, the  values  are the  mean $\aks$ over an  area of $10 \times 10$ arcmin  around the target position and based  on the \citet{Cardelli1989ApJ.345.245C} extinction law.   For the  bulge  area,  the total  extinction  $A_{Ks}$ was  taken directly from  the Bulge Extinction And  Metallicity (BEAM) Calculator\footnote{\url{http://mill.astro.puc.cl/BEAM/calculator.php}} \citep{Gonzalez2012AA.543A.13G} , and for  the disk  area, the  $A_{Ks}$ values    were   calculated    from   the EJK    map   presented    in \cite{Minniti2018AA...616A..26M}, assuming the \citet{Cardelli1989ApJ.345.245C}  law.  Relative  extinction for  the other VISTA filters  were provided by \citet{Catelan2011rrls.conf.145C} and are listed in Table \ref{tab:ext}  for the optical V band as well.

\begin{table}
\begin{center}
\caption{Relative extinctions  for the  $ZYJHK_{\rm s}$  VISTA filters,
  and      for       the optical      V      band.       Adapted      from
  \citet{Catelan2011rrls.conf.145C}.}
\label{tab:ext}
\begin{tabular}{c c}
\hline
\hline
\noalign{\smallskip}
Relative &  Value     \\
extinction &  \\
\noalign{\smallskip}
\hline
\noalign{\smallskip}
$A_{V}/A_{Ks}$  &  8.474 \\
$A_{Z}/A_{Ks}$  &  4.229 \\
$A_{Y}/A_{Ks}$  &  3.306 \\
$A_{J}/A_{Ks}$  &  2.373 \\
$A_{H}/A_{Ks}$  &  1.599 \\
\noalign{\smallskip}
\hline
\end{tabular}
\end{center}
\end{table}

\begin{figure*}
\centering
\includegraphics[angle=0,width=\textwidth]{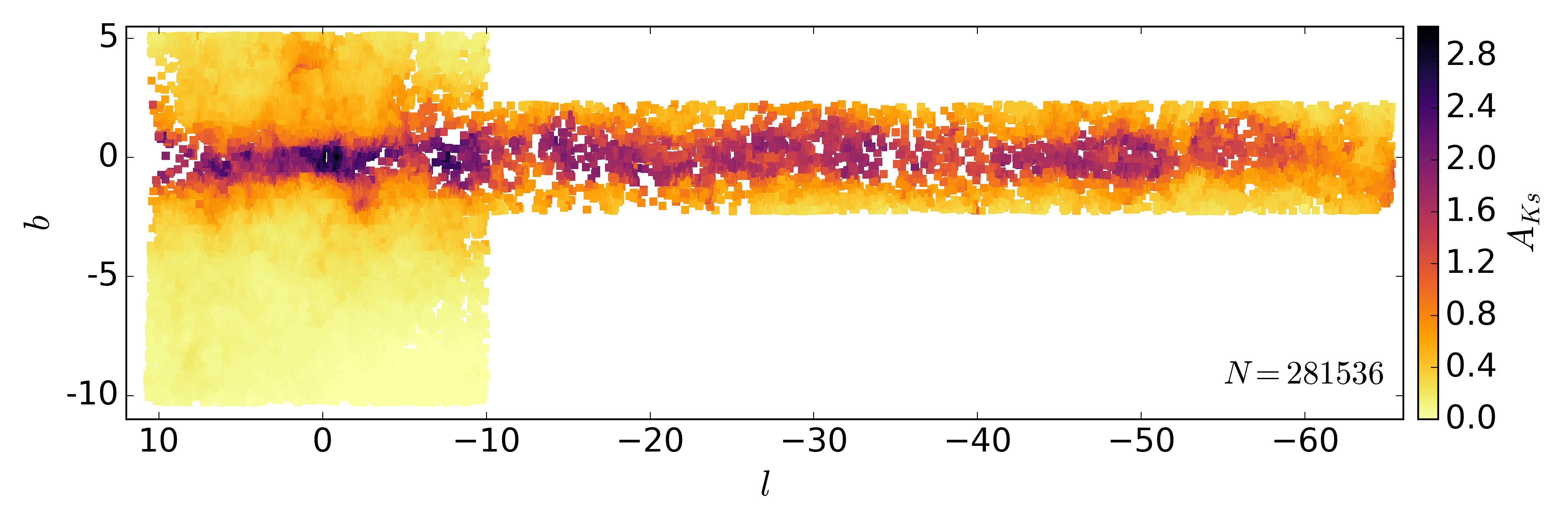}
\includegraphics[angle=0,width=\textwidth]{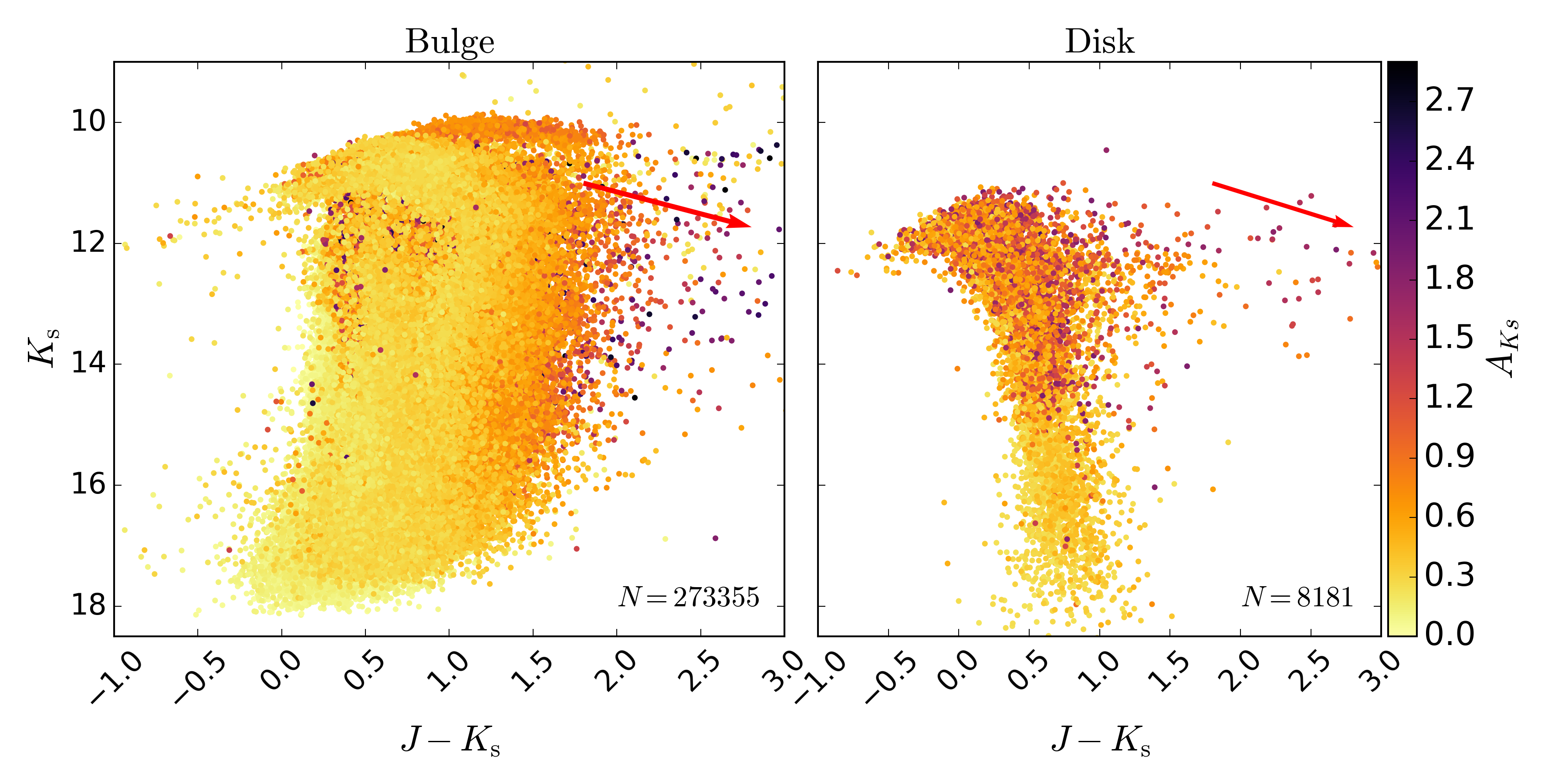}
\caption{ Top: Distribution for all 281\,536 stars in the
  match VVV+VSX. The data points are colour-coded by the
  total extinction \aks, integrated along the entire line of sight
  for each object. Values vary from $\aks < 0.01$~mag to $\aks
  \sim 3$~mag near the Galactic centre.  For nearby objects the total
  extinction as calculated by the VVV maps is certainly
  overestimated. Bottom: CMD diagram for the bulge (left) and disk (right) containing the variable stars of our sample colour-coded by the \aks parameter. The arrow shows the reddening vector for the diagram.}
\label{fig:AKs}
\end{figure*}

Fig.  \ref{fig:AKs} shows the surface density distribution along with the Galactic coordinates and the CMD for the disk and bulge for the V$^4$SX sample, colour-coded by the total extinction (integrated along the entire line of sight for each object) calculated for the VVV data \aks. \aks values for all objects are available electronically as supplementary material. \aks varies from $\aks < 0.01$~mag in  the outer bulge up to $\aks \sim 3$~mag  for objects  near the  Galactic  centre.  We note that for nearby objects in the foreground disk, the total extinction as calculated by the VVV maps is certainly overestimated.

\subsection{Cross-matching the VVV-VSX catalogue with Gaia-DR2}\label{sec:matchgaia}

Recently, another huge database has been publicly released. It is the second data release (DR2) of the optical Gaia mission, which contains positional data for over one billion sources at $G$, $BP,$ and $RP$ bands, as well as precise parallaxes and proper motion measurements \citep{Gaia2018AA...616A...1G}.  The dataset allowed us to match the VVV+VSX  and  Gaia-DR2  catalogues, selecting the matches with the smaller separation within 1 arcsec radius. A total of 590\,824 pairs were found within 1 arcsec, for which parallaxes are available in  Gaia DR2 (we did not distinguish between positive and negative parallaxes because of the biases that this procedure would introduce in the sample, e.g. \citealt{Bailer-Jones2018AJ....156...58B}, BJ18 hereafter), and we can estimate the distance to them assuming a naive value of $d=1/\omega$. Distances based on  Gaia parallaxes were also retrieved from \citetalias{Bailer-Jones2018AJ....156...58B}, obtained on the basis of a three-dimensional model of the  Galaxy instead of those obtained by simply inverting the parallax. From the total initial variables in our sample, we found 207\,439 pairs for which BJ18 distances are available. A comprehensive description of distance estimates based on Gaia parallaxes is presented in \citet{2018AA...616A...9L}. The  full description of all  columns from Gaia that appear in the catalogue is given in Section \ref{sec:coldesc}.

\begin{figure*}[!]
\centering
\includegraphics[angle=0,width=\textwidth]{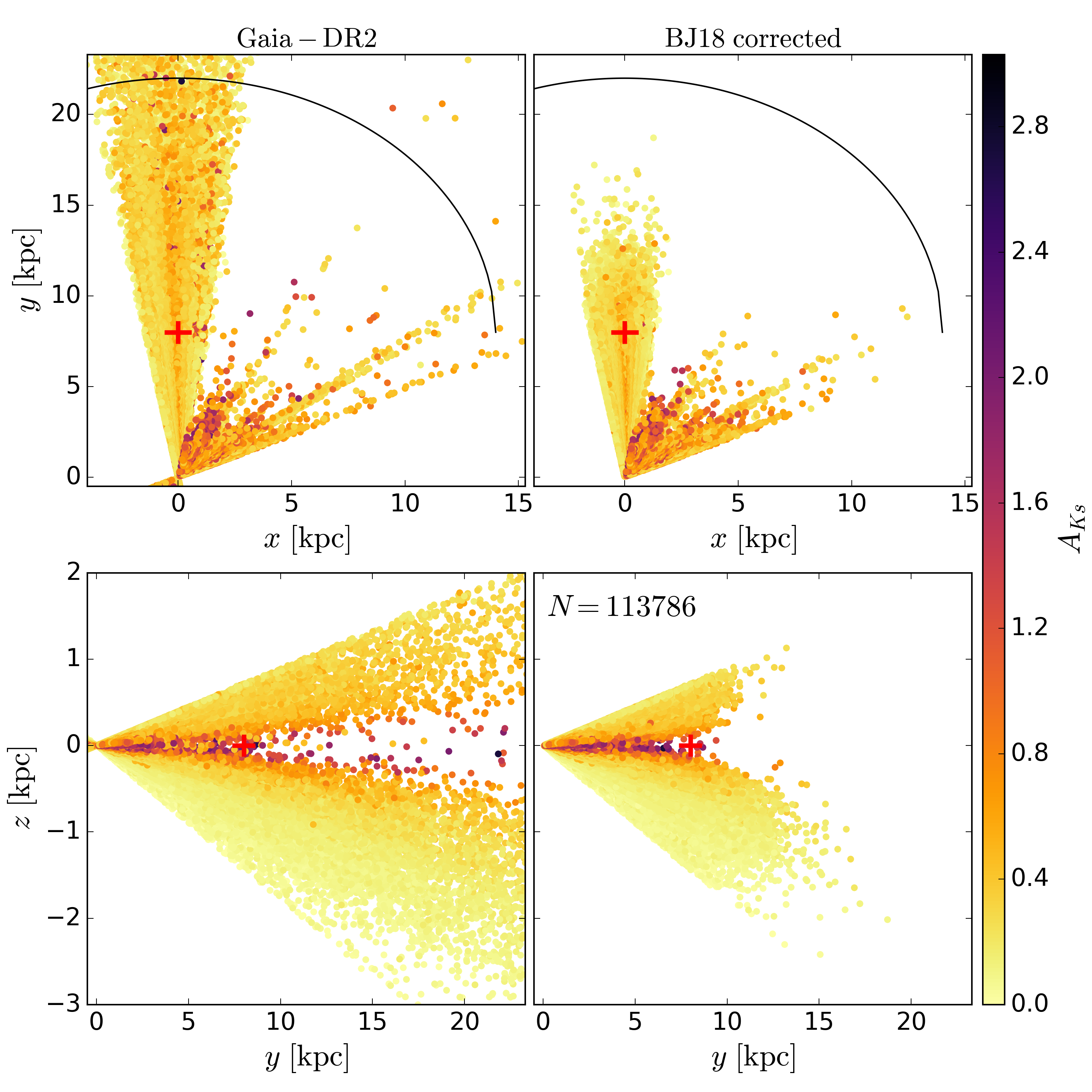}
\caption{Distances calculated directly from the Gaia parallax (left)
  and   distances  from   \citetalias{Bailer-Jones2018AJ....156...58B}
  (right).  Upper  panels: $(x,y)$ projection. The  black semi-circle
  identifies the 15\,kpc radius from the Galactic centre as  a symbolic limit
  for the  disk of the  Galaxy. Lower panels: $(y,z)$  projection.   The  red   cross  identifies  the  geometrical  position adopted  to the centre  of the Galaxy  assuming $d=8$\,kpc. The points are colour-coded by the total extinction in the \ks band. The sample is constrained in the VVV $J$ and $K$s magnitudes and for those stars with distances measured using \citetalias{Bailer-Jones2018AJ....156...58B}.}
\label{fig:gaia}
\end{figure*}

Fig.   \ref{fig:gaia}  shows  these resulting sources coloured by the \aks as adopted for Fig. \ref{fig:AKs}, but  for  the distances  calculated  from  the Gaia  parallax. Selecting only VVV sources whose $J$ and $K$s magnitudes have flags between -9 and 0 (see Section \ref{sec:match} for a more detailed overview of the meaning of the flags) and with the distance measured by \citetalias{Bailer-Jones2018AJ....156...58B}, we obtain a sample containing 113\,786 objects. The data  we release contain  all VVV+VSX
matched  variables, regardless  of  whether they are  good according to  the $JK_{\rm s}$ flag criteria.  The discrepancy between  the measured  distance directly from  the parallax and that from  taken from \citetalias{Bailer-Jones2018AJ....156...58B} is quite evident. We can fairly assume that this correction is highly important because many  of  the Galactic  sources in  our sample would be located much farther away than they should when only considering the parallax, i.e. Galactic sources have distances bigger than the size of the Milk Way. Fig. \ref{fig:gaia} also shows the projected distribution of stars  along with the height  $z$  of  the Galaxy  as  a  function of  the geometrical distance  $y$. It  is interesting to  note that  the stars appear to avoid the Galactic centre (around $b = 0^\circ$), as is also visible in the spatial distribution of Fig. \ref{fig:AKs}. This is a natural  and known phenomenon and is the consequence of the high interstellar extinction along the line of sight of the disk and bulge.

\subsection{Column description}\label{sec:coldesc}

As Table  \ref{tab:matchedtable} does not show many of the columns in our catalogue, in this section we describe all the columns from VVV that were modified or created by us that are available electronically as supplementary material (we do not describe those of the already public catalogues such as VSX and Gaia).

 The following columns are directly obtained from the VVV catalogue:
 
\begin{itemize}
\item   \texttt{RA\_(VVV)}   and    \texttt{DEC\_(VVV):}      J2000.0
  coordinates from the VVV catalogue (in degrees).

\item \texttt{L}  and \texttt{B:}   The Galactic coordinates  $l$ and
  $b$ as obtained for the VVV (in degrees).

\item \texttt{MAG\_}\textit{i:} Magnitude of the sources measured by
  VVV,  where $i  = \left\{Z,  Y, J,  H, K\mathrm{s}\right\}$  are the
  respective bands of the VVV survey.

\item \texttt{ERR\_}\textit{i:} Errors of the magnitudes  of VVV to
  the respective bands $i = \left\{Z, Y, J, H, K\mathrm{s}\right\}$.

\item  \texttt{F\_}\textit{i:} The photometric flags for the bands where $i = \left\{Z, Y, J, H, \ks\right\}$.

\end{itemize}

The following columns are obtained from the information of VVV catalogue or from the complete match VVV+VSX+Gaia:

\begin{itemize}

\item  \texttt{$A_{Ks}$}  : The  total  extinction  in  the  \ks\  band.

\item  \texttt{distance:}  Distances calculated  using  the  Gaia
  parallax ($d=1/\omega$, in kpc).

\item  \texttt{dist\_BJ:} Distances  from \citetalias{Bailer-Jones2018AJ....156...58B} (estimated distance `r\_est', in kpc).
  
\item  \texttt{(x,y,z)\_BJ:}  Distances  based on \texttt{dist\_BJ}.

\item \texttt{CandidateType:}  Variables reclassified in Section~\ref{sec:varGaia}.

\end{itemize}

%\section{The VSX classes of variables through diagrams}\label{sec:discussion}

%\subsection{The variables in VVV separated by VSX classes}\label{sec:vsxclasses}

\begin{figure*}[!]
\centering
\includegraphics[angle=0,width=\textwidth]{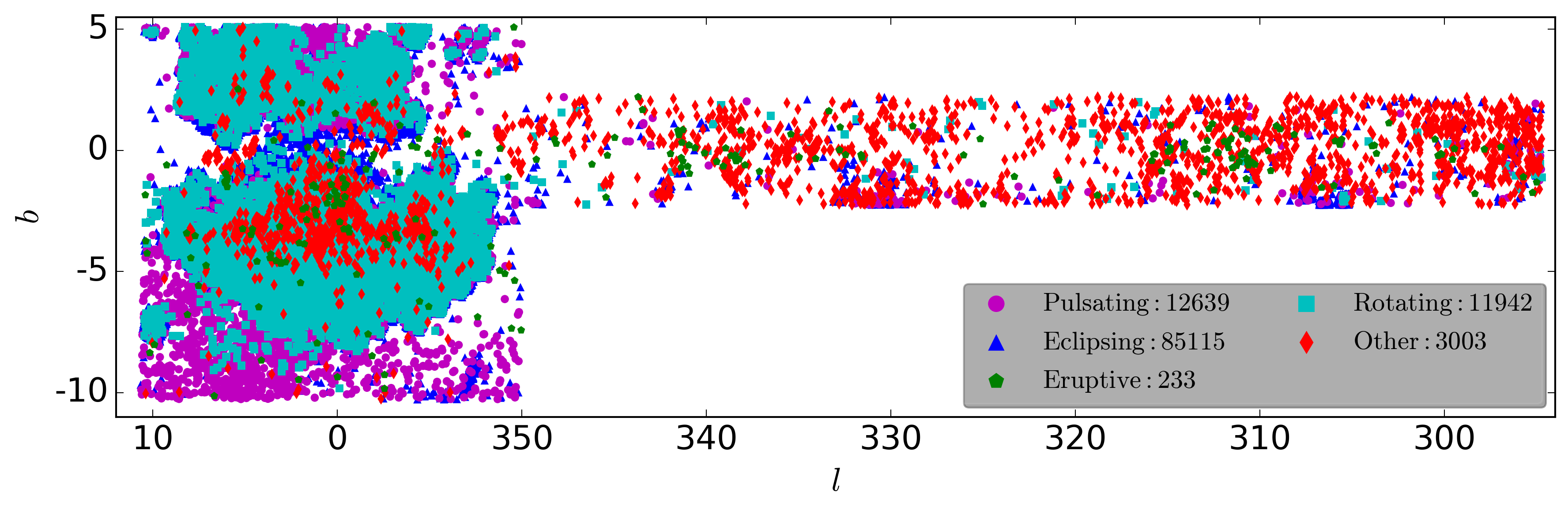}
\caption{All 112\,932 variables of our sample distributed through the five main classes we defined with the Galactic coordinates $l$ and $b$. Symbols are colour-coded following the legend in the figure.}
\label{fig:allvars}
\end{figure*}

The  surface density distribution presented  by the Fig. \ref{fig:allvars}  is irregular and allows different interpretations in terms of stellar population and Galactic structure. For instance, a smaller number  of objects is seen in the  innermost bulge area.  This
so-called zone of avoidance is also present in the distribution of the VVV
Novae catalogue  \citep{Saito2013AA.554A.123S} and is caused by  the high
extinction that affects the observations in  the optical wavelengths,
as seen in most variability  surveys. This absence of objects in the innermost regions is even
more evident  in the  distribution of long-period variable candidates
from   Gaia  \citep{Mowlavi2018AA...618A..58M},   which is also   based  on   optical
observations.   Observable  objects  in  these regions  are  mainly from  the
foreground disk,  as can  be seen in  the $y,z$  projected distances
(bottom panels  of Fig.  \ref{fig:gaia}).   On the other  hand, closer
objects are  probably saturated in  the VVV observations and thus  are not
included in our sample.

The highest density of variable sources is seen in the intermediate bulge region  ($-10\degree < b -5\degree$) of Fig. \ref{fig:allvars} and is mostly caused by  RR Lyrae that were detected in variability surveys such as  OGLE and VVV.   Especially for  OGLE, the footprint is easily seen across the bulge \citep[see   figure   7   in][]{2011AcA....61....1S}. RR  Lyrae are metal-poor  population II stars that  are detected in large numbers  in the bulge region, compared to a small fraction that is present throughout the Galactic disk, which  should   be  dominated  by   metal-rich  population I stars \citep[e.g.][]{2018ApJ...857...54D,2020arXiv200802280I}.

\subsection{Some statistics}\label{sec:othervvv}

This section is dedicated to gathering the information available for our VVV+VSX+Gaia  catalogue.  From the 701\,256 variables in the match VSX+VVV, we selected all objects with flags lower than 0 and greater than -9 ( i.e. unsaturated point sources).   We have the following numbers: 474\,105 in $Z$ band, 420\,775 in $Y$, 366\,282 in $J$, 353\,220 in $H$, and 346\,614  in $K_{\rm s}$. There are 195\,580 sources with good photometric quality in all five bands.

From the 590\,824 objects with Gaia parallaxes,  206\,624 have negative measures, with a minimum value of -443.735. These parallaxes are unreliable and hence were not used for any analysis involving distances, although the stars are not excluded from the table. The highest parallax value is for an  M-type variable star in the bulge  direction: 282.189 arcsec.   For the positive  values of the parallax,  the mean value  is  0.468\,arcsec  with a  standard  deviation  ($\sigma$)  of 0.990.  The mean  corresponding  distance from  the naive  Gaia's parallax  is 28.326\,kpc   with  $\sigma = 1711.582$\,kpc. For the \citetalias{Bailer-Jones2018AJ....156...58B}  distances, the average is $\overline{d_{\rm BJ18}} = 3.246$\,kpc with $\sigma = 1.625$\,kpc.

\subsubsection{Analysis of the VVV photo-flags}\label{sec:flags}

\begin{figure*}[!]
\centering
\includegraphics[width=\textwidth]{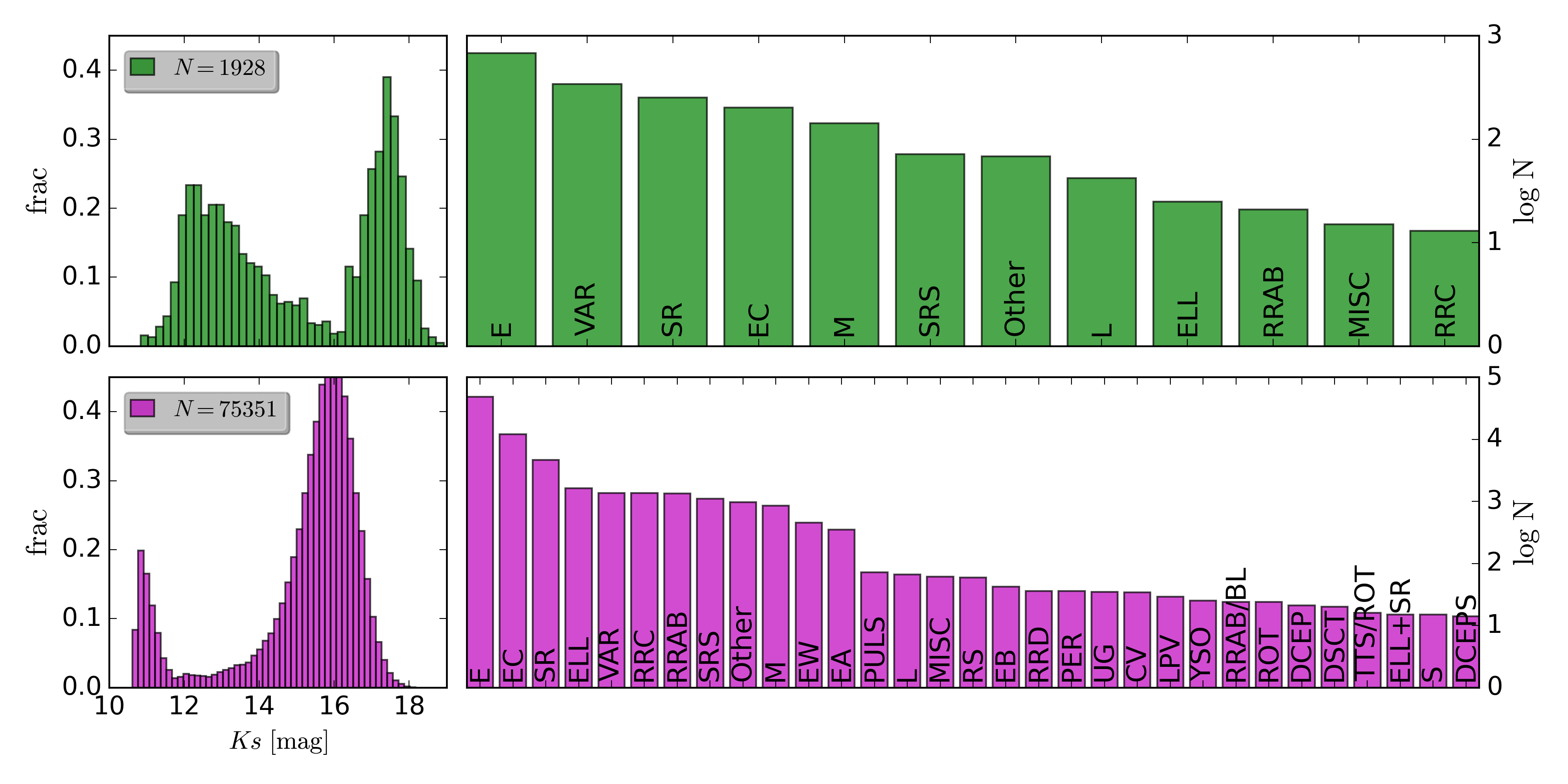}
\caption{Green: Histogram (left) for the sources with flag 0 (noise) for the \ks band and their distributions in the VSX classes (right side). Magenta: Same, but for flag 1 (non-stellar objects) for the \ks band.}
\label{fig:flags}
\end{figure*}

During the definition of the sample for the analysis of the VIVA catalogue, we realised that a considerable number of stars have flags 0 (noise) and 1 (non-stellar objects), but most of them are compatible and classified as point sources in VSX. We analysed the conditions of these measurements and briefly investigated whether they are misclassifications or spurious data. Fig.~\ref{fig:flags} shows the histograms of the stars (as identified in VSX) collected for these two flags. Out of 701\,256, 2393 (0.34\%) that were flagged as noise ($F\_\ks = 0$) and are shown in the green histogram in Fig.~\ref{fig:flags}, about half of them are fainter than the detection limit (the flag for the \ks band is typically 17.5, although not constant, depending on the extinction of the region being considered)\footnote{\url{http://www.eso.org/rm/api/v1/public/releaseDescriptions/80}}. The right side of Fig.~\ref{fig:flags} shows the number of variables corresponding to each defined VSX class. For this analysis, all objects assigned as suspected variables in VSX (for which a colon is added to the class) were added to the group ``Other''.

While flag 0 contains just a small number of objects, negligible for most purposes, flag 1 has a much more significant number of stars. We found that 133\,808 (19\%) stars in our main sample are classified as non-stellar objects. Fig.~\ref{fig:flags} shows the histogram and the distribution per variable class for these objects, and although part of the sample lies beyond (or very close to) the detection limit, these sources are better behaved, and most of them are within the saturation-detection limits. Moreover, most of the sources ($\sim 77$\%) with $F\_Ks = 1$ are eclipsing variables, with RR Lyrae and Mira following as second and third most common types in the sample.

We verified a similar behaviour for the other VVV bands \textit{ZYJH} and find the same conclusions with similar numbers. This information can be further used to constrain their definitions and maybe improve the accuracy rate of these flags for future releases of the extended VVV survey (VVVX\footnote{\url{https://www.eso.org/sci/observing/PublicSurveys/docs/VVVX_SMP_07022017.pdf}}).

\section{Variables through distance and period}\label{sec:varGaia}

As the number of different types of variables is high and thus prevents a clearer view in the diagrams, the similar variable types were merged following the VSX\footnote{ \url{https://www.aavso.org/vsx/index.php?view=about.vartypes}}  denomination for simplicity.  We separated them into the five main groups following the definitions of the VSX  for Eruptive, Rotating, Cataclysmic, Pulsating and Other, which contains all objects that are not included in the previous groups. The last group alone does not follow the VSX  denomination strictly and might contain a mix of stars with unknown variability type and/or exotic sources such as active galactic nuclei, gama-ray bursts, quasars, and microlensing events. We did not add the composite types from VSX to these group (e.g. Algol systems, in which the low-mass component is close to its inner Roche lobe,  EA/SD ), for instance.

\begin{figure*}[!]
\centering
\includegraphics[width=\textwidth]{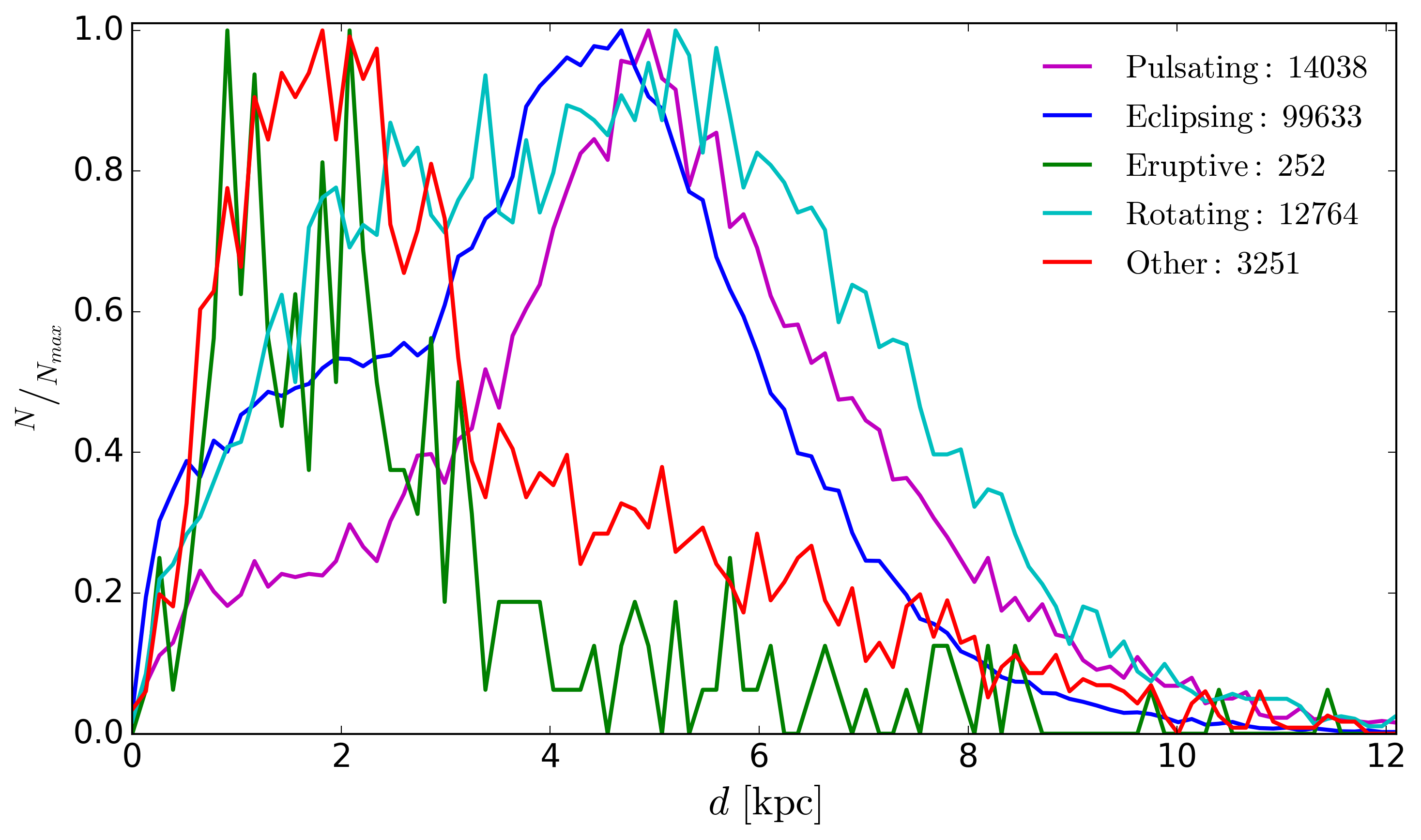}
\caption{Histograms for the VSX classes of variables normalized to 1 relative to the distance calculated as in the Section \ref{sec:matchgaia}.}
\label{fig:hist_classes}
\end{figure*}

We here then present some results using the period from VSX and other properties in terms of the five groups we just defined. For instance,  we only used the \ks magnitude and selected the good sources accordingly using only the flags in this band to constrain our sample. As a result, we have a higher number of variables, with a total of 130\,855 objects, of which we show 129\,938 in Fig. \ref{fig:hist_classes}. The missing objects are stars that we were unable to classify into any of the five main classes defined here.

Fig. \ref{fig:hist_classes} shows the histograms normalised to 1 at their maximum for the defined classes of variables as a function of the distance calculated using the \citetalias{Bailer-Jones2018AJ....156...58B} distances. The Eruptive stars have two peaks at 1.3~kpc and 2.1~kpc, while the Other group exhibits a broad distribution from $\sim$0.5~kpc to $\sim$3~kpc and a tail extending up to 10~kpc. As this group contains different types of variables, this is an expected effect, possibly composed of a mix of Pulsating and Rotating stars. This assumption is being explored by other works.

Because exploring all the VSX classes, which are fundamentally related to our five main groups, is beyond of the scope of this work, we concentrated on one large group, the Pulsating stars, and three particular VSX types from the group Other. They are the VSX types MISC (stars whose classification was not precise enough with the automatic methods that were applied and they can be either red variables or irregular types stars), S (variables with rapid light changes that have not been studied so far), and VAR (unclassified variable stars).

\begin{figure}[!]
\centering
\includegraphics[width=\columnwidth]{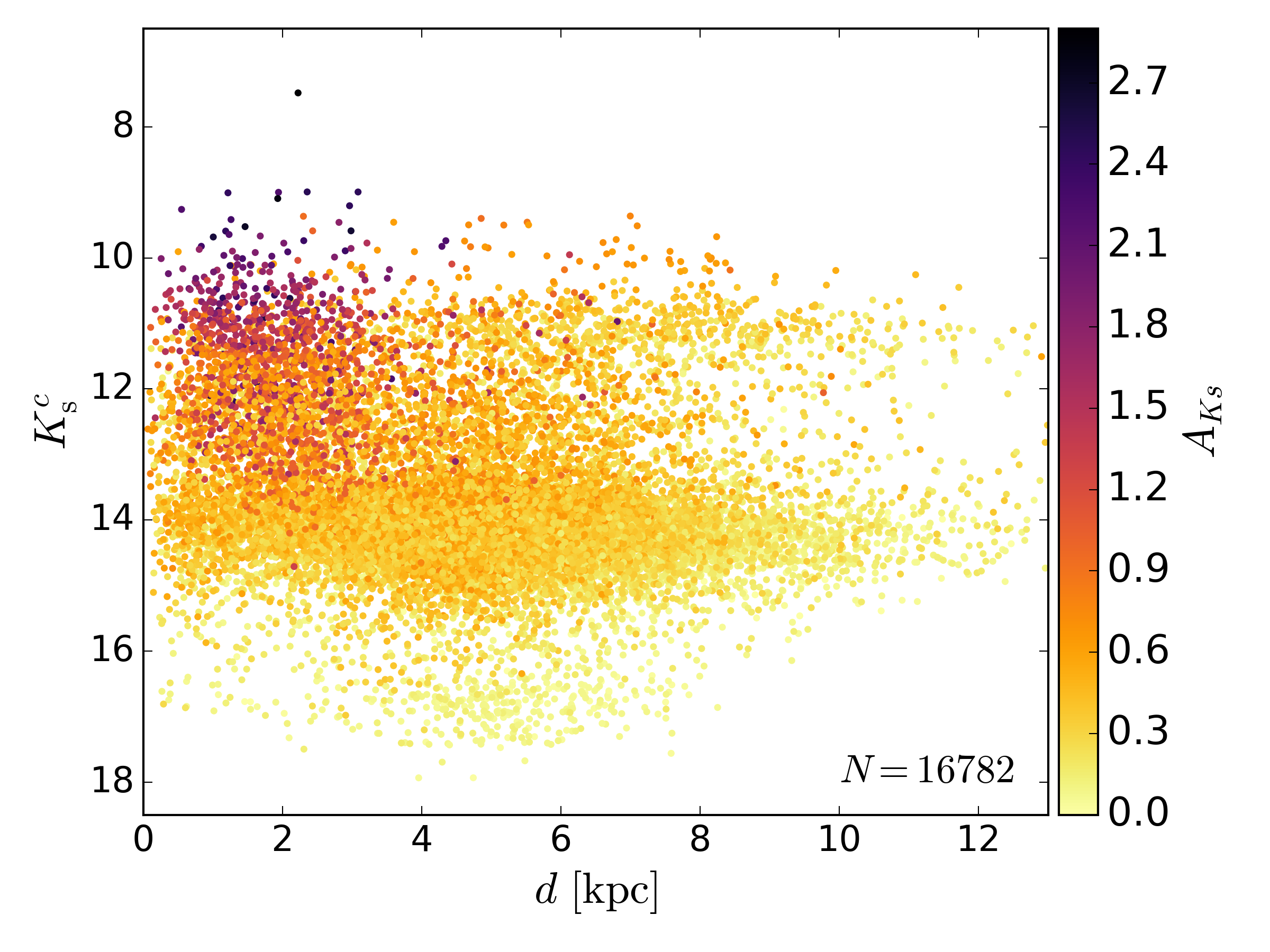}
\caption{Distance of the group defined as Pulsating stars along with the VSX types MISC, S, UNC, and VAR from the group Other as a function of the extinction-corrected magnitude \ks colour-coded by extinction \aks.}
\label{fig:distance_magnitude}
\end{figure}

The result of this selection are the 167\,892 stars (here we introduce another cut to the sample because of the \aks measurements, which cause 507 stars to be dropped from consideration)  shown in Fig. \ref{fig:distance_magnitude} for the extinction-corrected \ks magnitude ($\ks^c$) as a function of the \citetalias{Bailer-Jones2018AJ....156...58B} distance and colour-coded by the extinction \aks. We can identify five main groups: four horizontal groups ($\ks^c > 16$, $\ks^c \sim 14.5$, $\ks^c \sim 12.5$, and $\ks^c < 12$) for all distance ranges, and the fifth group, which spans from $9 < \ks^c \lesssim 14$ but is concentrated at smaller distances ($d < 3$~kpc).

\begin{figure}[!]
\centering
\includegraphics[width=\columnwidth]{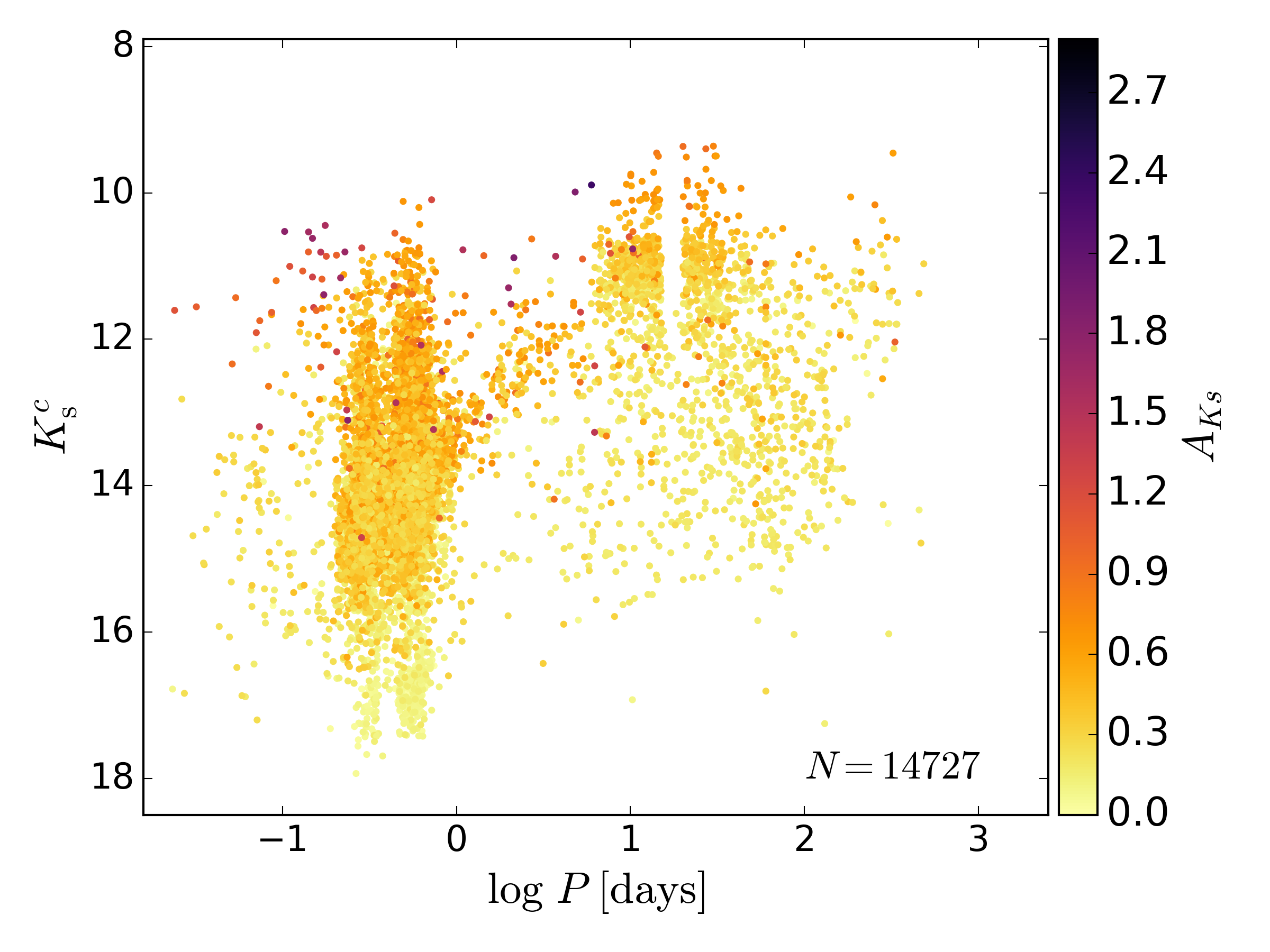}
\caption{Extinction-corrected \ks magnitude as a function of the period from the VSX catalogue, colour-coded by \aks.}
\label{fig:period_MAG_Ks}
\end{figure}

In the same manner, Fig. \ref{fig:period_MAG_Ks} shows the diagram of the period retrieved from the VSX catalogue as a function of the extinction-corrected \ks magnitude for all variables shown in Fig. \ref{fig:distance_magnitude}, except for those that do not have a measured period in the VSX catalogue for a total of 14\,727 objects. The 2055 objects that are missing in this sample relative to those shown by Fig.
~\ref{fig:distance_magnitude} correspond to the stars with undetermined periods, and Fig. \ref{fig:distance_magnitude} and Fig. \ref{fig:period_MAG_Ks} show that they are mostly objects of high extinction that are mainly located at the centre of Galactic plane (see Fig. \ref{fig:AKs}). Fig. \ref{fig:period_MAG_Ks} shows two dominant groups, most composed of RR Lyrae (shorter periods) and Cepheid stars (longer periods). We study the VSX classes individually in this projection.

\begin{figure*}[!]
\centering
\includegraphics[width=\textwidth]{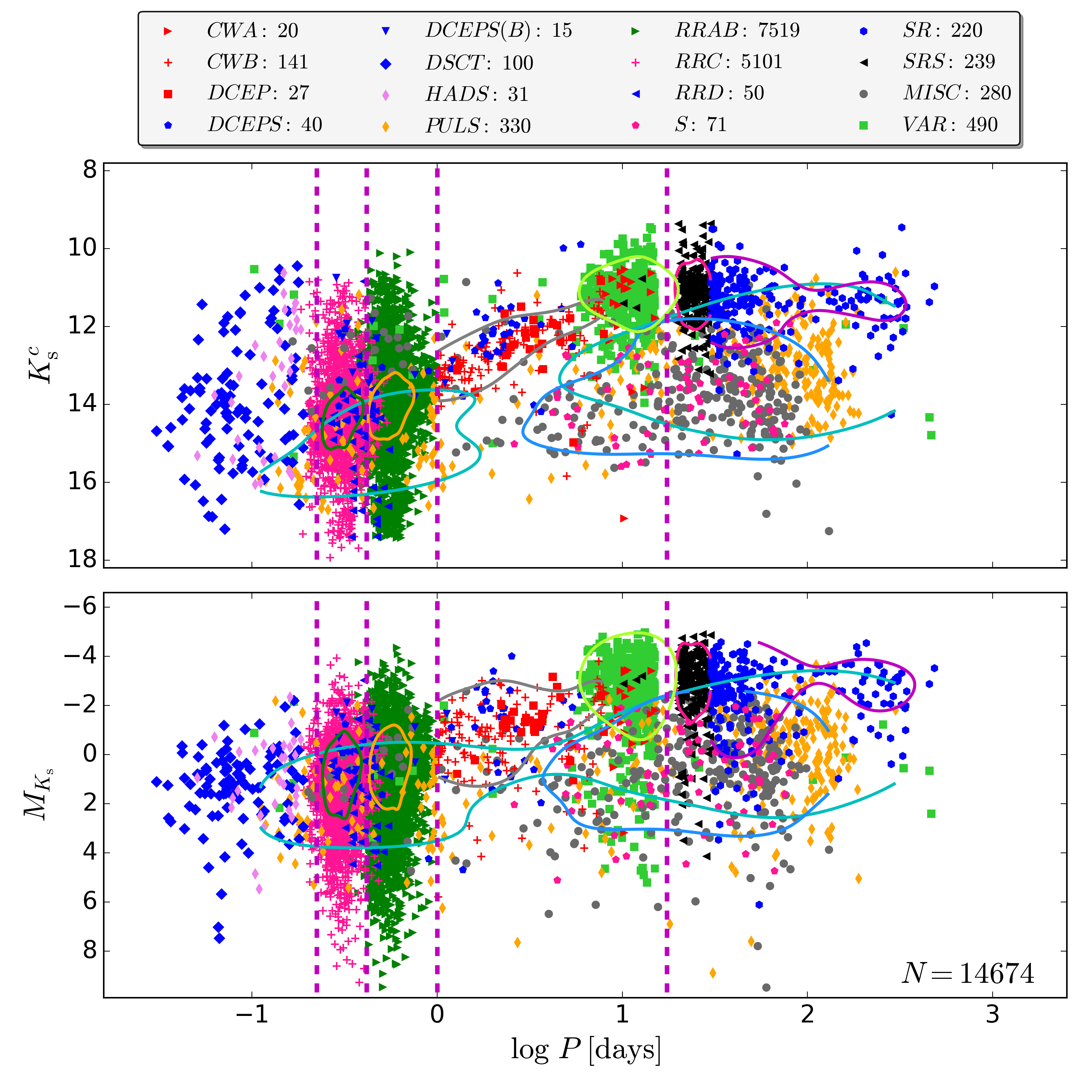}
\caption{Diagram for the period as a function of the extinction-corrected \ks (top panel) and absolute $M_{\ks}$ (bottom panel) magnitudes. The VSX variable types are highlighted. VSX types with fewer than ten stars were removed because they are not statistically relevant. The group called LPV (long-period variables) was removed because the measurement of its period in the VSX catalogue has low precision. Contours represent the 68th percentile (the contour colour differs from the point colour) of the types with more than 100 points and are represented as RRAB by the yellow contour, RRC in dark green, PULS in cyan, CWB in gray, MISC in blue, SR by the purple line, and VAR by the golden line. The vertical dashed purple lines represent the separation limits between DSCT and RRC ($\log\ P = -0.65$), RRC and RRAB ($\log\ P = -0.38$), above RRAB ($\log\ P = 0$), and SR from short-period variables ($\log\ P = 1.24$). The number $N$ in the bottom panel is the total number of variables shown.}
\label{fig:period_class_MAG_Ks}
\end{figure*}

To extend the study of Pulsating stars and the VSX variables types, we plot them in a diagram of magnitude \ks (the apparent and the absolute) as a function of the period, but identifying the variable types that are considered (see Appendix \ref{sec:appenB} for the full names and descriptions of each type in Fig. \ref{fig:period_class_MAG_Ks}). The results are shown in Fig. \ref{fig:period_class_MAG_Ks} along with the contours representing the $1\,\sigma$ boundary for the more populous groups. We easily identify two main distinct regions populated by RR Lyrae on one side and the semi-regular variables on the other. In between these two main groups lie the W Virginis variables (CWA and CWB), which occupy the upper portion of the diagram and several unstudied variables that occupy the lower portion (e.g. the unstudied variables with rapid light changes, S, form a well-defined group in this region along with other non-classified variables of the types MISC, PULS, and VAR).

The apparent magnitude of the types shown in Fig. \ref{fig:period_class_MAG_Ks}  appears to be very well-defined in different regions of the diagram, they become more inhomogeneous when distributed over the period versus absolute magnitude plane, at least on the magnitude axis. We can still draw vertical lines in both panels of Fig. \ref{fig:period_class_MAG_Ks} using the contours as parameters for the divisor line. The first line is drawn at $\log\ P = -0.65$ and separates the $\delta$ Scuti type of variables (DSCT and HADS) from the RR Lyrae with symmetric light curves (RRC). The second line is located at $\log\ P = -0.38$ and separates the two RR Lyrae groups RRC and RRAB, the second being the RR Lyrae with asymmetric light curves. The third stands for all variables with higher periods than the RRAB and is positioned at $\log\ P = 0$. Finally, the fourth line is drawn at $\log\ P = 1.24$ and separates the semi-regular variables (SR) from those with shorter periods.

Based on these definitions, we can now identify the unclassified stars from our sample into the most populous VSX types, at least as a first approximation. We can  then classify the types called MISC, PULS, and VAR depending upon which side of the lines they occupy. We did not classify type S when they occupy a region of the diagram that is not populated by any other well-defined VSX type in our sample (at $0 < \log\ d < 1.24$ and just below where the CWA, CWB, and SRS are located). They might be W Virginis or semi-regular pulsating variables or even more probably, they are variables other than pulsating stars and are to be classified accurately through other observations. We considered type S as a variability type of stars and included MISC, PULS, and VAR located at $0 < \log\ d < 1.24$ as S types. Our classification is loosely based on about two parameters: period and \ks magnitude, and the term can be misread as a determination when in fact we intend to provide candidates that need to be studied in more detail to confirm their final type. Applying the definitions above (for the entire table) with good \ks photometry and well-determined periods, we account for a total of 337\,559 stars that are available for a reclassification, we find the number of candidates listed in Table \ref{tab:classific}. No PULS were considered. These results are included in the supplementary material as a new column in the table called \texttt{CandidateType}. We call attention to the stars of VAR type, however, which are by definition variables of undefined type. This nomenclature is also used for variable candidates, which makes the period found in the VSX database not completely reliable and implies that the strange gap around $\log\ P = 1.24$ might be an artefact due to incorrect estimates of the period (we thank the anonymous referee for pointing this out). As a consequence, we must be careful when these stars are used for science. On the other hand, we consider them important objects to be targeted by variability surveys and/or follow-ups, therefore we kept them in the analysis of Fig. \ref{fig:period_class_MAG_Ks}.

\begin{table}[!]
\begin{center}
\caption{Candidates to VSX variable types for the unclassified classes MISC, PULS, and VAR using the period vs. \ks magnitude space from Fig. \ref{fig:period_class_MAG_Ks}.}
\label{tab:classific}
\begin{tabular}{ccccccc}
\hline
\hline
\noalign{\smallskip}
Type & DSCT & RRC & RRAB & S & SR \\
\noalign{\smallskip}
\hline
\noalign{\smallskip}
MISC  & 17 & 22 & 27 & 122 & 371\\
VAR  & 5 & 1 & 6 & 23\,681 & 24 \\
\noalign{\smallskip}
\hline
\end{tabular}
\end{center}
\end{table}

\section{VIVA comparison}\label{sec:viva}

 The VIVA catalogue \citep[][]{FerreiraLopes2020MNRAS.496.1730F} is based on a variability analysis of the VVV-DR4 Data Release that is the same as we used here. It was compiled following a series of recommendations in the New Insight into Time Series Analysis \citep[NITSA - ][]{FerreiraLopes2016A&A...586A..36F,FerreiraLopes2017A&A...604A.121F,FerreiraLopes2018MNRAS.481.3083F}. These criteria provide a good prescription for selecting variable stars. We caution that the VIVA catalogue only considered stars with more than ten non-flagged observations (number of good observations, or $\texttt{NG} > 10$), that is, detection-flag (\texttt{ppErBits}\footnote{The full description of the detection flag can be found at \url{http://horus.roe.ac.uk/vsa/ppErrBits.html}.}) smaller than 256 and stack-flag (\texttt{flag}) equal to 0. The first is related to the archive curation procedures and the second to potential matching problems. These thresholds were adopted to reduce problems regarding the integrity of the detection. On the other hand, all VVV sources having at least one observation and at least one counterpart with VSX catalogue within an 1-arcsec radius are considered in V$^4$SX.
 
By matching with V$^4$SX within a radius of 1 arcsec, we found that 330838 single-detection point sources are not present in the VIVA catalogue. Three main reasons might be the origin of these differences: (a) our sample contains sources with fewer than ten non-flagged observations, (b) the signals reported in other wavelengths are not clear in the IR, and (c) the procedures used by \citet{FerreiraLopes2020MNRAS.496.1730F} fail. To understand these differences, we analysed the stars whose periods are reported in the literature, using the signal-to-noise ratio (S/N) estimated with harmonic fits to select our sample \citep[e.g. ][]{FerreiraLopes2015A&A...573A.100F,FerreiraLopes2015A&A...583A.122F}.

\begin{figure*}[!]
\centering

\includegraphics[width=1\textwidth]{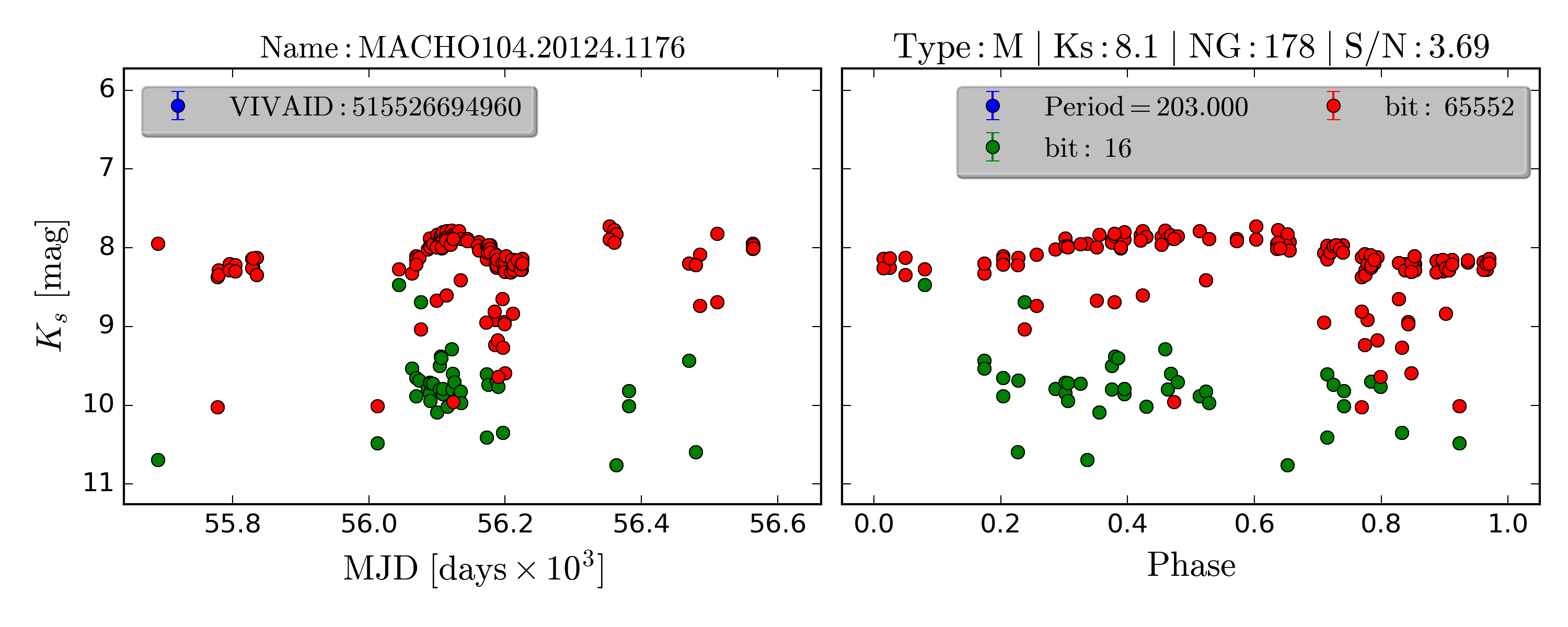}%a

\includegraphics[width=1\textwidth]{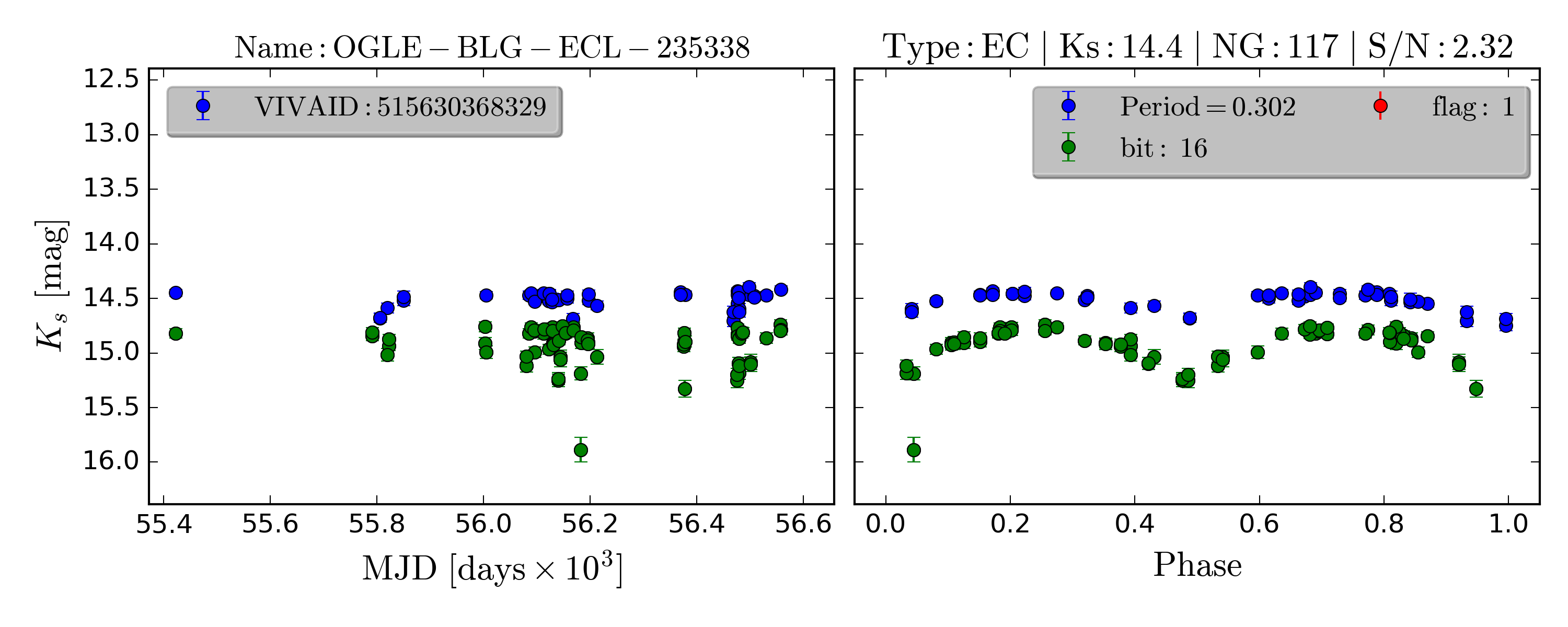}%e

\includegraphics[width=1\textwidth]{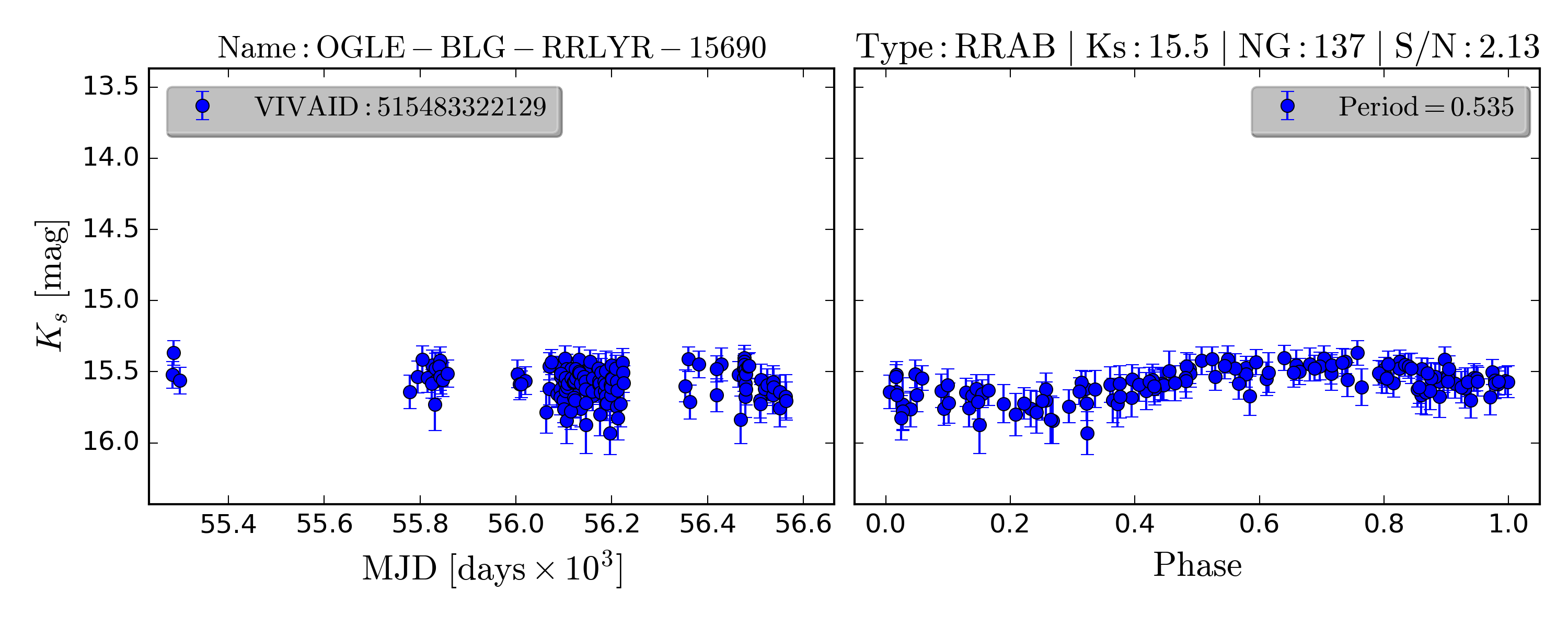}%h

\caption{Time domain (on the left side) and phased light curves (on the right side) for examples of variable stars that are missing in the VIVA catalogue. The light curves display the VSX nameat the top and the VIVA ID inside the legend box, and the corresponding phase diagram shows the VSX type (Type), the VVV \ks magnitude, the number of good observations \texttt{NG,} and the S/N estimated with harmonic fits at the top, with the VSX period and the \texttt{ppErrbits} flags in the legend box inside the diagram. Blue points represent $\texttt{ppErrbits} = 0$ for good observations, and red and green stand for observations with $\texttt{ppErrbits} \neq 0$.}
\label{fig:lcs_fiva}
\end{figure*}

We selected a sample of 10\,793 stars considering two criteria: $\mathrm{S/N} > 3$ and a number of good observations exceeding $20$ or $\mathrm{S/N} > 2$ and \texttt{NG} higher than $100$. Following the VSX classification, this sample is composed mainly of eclipsing binaries ($\sim63.5\%$), semi-regular variable stars ($\sim23.5\%$), variables of unknown type ($\sim8.4\%$), and RR Lyrae type stars ($\sim3.2\%$). From these, we selected 1\,195 sources by visual inspection of the phase diagram and light curve (side by side, as shown in Fig. \ref{fig:lcs_fiva}). The flagged observations were also displayed to verify their weight on the data. During this process, we observed many objects that appeared to be similar to Algol-type stars, whose periods might be inaccurate \citep[][]{Carmo2020MNRAS.498.2833C}, as well as some that appeared to have a very low amplitude signal.

On the other hand, for the selected objects displaying good signals (for which three examples are shown in Fig. \ref{fig:lcs_fiva}), we observed that $\sim 98\%$ of them belong to the VVV bulge fields. These regions are more crowded and dustier than those located in the disk area, and as a consequence,  the number of missed sources is expected to be higher. Additionally, noting that the peak of the magnitude distribution of our sample is at $\ks \sim 15$ mags, the missed sources are in general faint objects with a low S/N (see middle panel of Fig.~\ref{fig:lcs_fiva}). These correspond to less than $0.5\%$ of the variable stars with more than ten non-flagged observations. We also observed that some flagged observations may be useful for identifying LPVs (see the top diagrams in Fig.~\ref{fig:lcs_fiva}). Only the LPVs appear to show variability at the flagged measurements. On the other hand, the flagged observations might also indicate the zero-point error (see the middle panels in Fig.~\ref{fig:lcs_fiva}).  We caution that these observations must be used carefully because they can also increase the noise of the signal.

In summary, our results agree with those of \citet{FerreiraLopes2020MNRAS.496.1730F}, that is, the VIVA catalogue includes about $99\%$ of all variable stars found in the VVV Data Release 4 that have good signal. The missed sources correspond to faint stars mainly with a low S/N. Additional observations from VVVX can help solve this matter. Moreover, \citet[][]{Hadju2020ExA....49..217H} identified two independent types of bias in the photometric zero-points in VVV data that can also improve the data quality.

\section{Conclusions}\label{sec:conclude}

We presented a  near-infrared catalogue that includes accurate individual coordinates, magnitudes, and extinctions as well as distances based on Gaia parallaxes for variable stars presented in the VSX catalogue and also in the VVV near-infrared colour catalogue.   Our  CMDs,  colour-colour diagrams, and surface density  distributions for the different types allow us to give a global characterization of Galactic variables stars, including important distance indicators such as RR  Lyrae, Cepheids, abd Miras.

Our results suggest that the current knowledge of variability in the Galaxy is biased to nearby low extinct stars and that therefore the larger part of the  Galaxy disk and bulge is still unexplored by stellar variability studies.

We provide a set of several thousand candidates to different VSX types of variables using the period space to constrain the region they occupy in the period versus magnitude diagram. This result can be used to target specific samples of stars to determine the characteristics through follow-up campaigns.

Future studies of the Milky Way variable stars will be revolutionised by the LSST survey in the next decade, for which the discovery of hundreds of thousands of variables is expected. Thus, the present catalogue also provides the groundwork to characterise the results of future projects. Another important aspect to consider is the near-infrared versus optical variability. The searches performed at different wavelengths result in different relative numbers of variables, and this must be considered when the number of Rubin Observatory findings in this field is predicted \citep[e.g. ][]{Pietrukowicz2012AA...537A.116P}.

Near-infrared surveys such as the VVV are efficient at low Galactic latitudes close to the plane, where optical surveys are usually blinded by the absorption in the interstellar medium. In the same way that Baade's Window was important for optical studies of the Galactic bulge, the near-infrared surveys also profit from the study of the windows that have recently been found in the Milky Way plane \citep{Minniti2018AA...616A..26M}.

By analysing the photometric flags from the VVV catalogue, we identified a misclassification regarding noise and mainly non-stellar objects. These stars are misclassified even inside the magnitude range that allows a good verification of the data and typically corresponded to $\sim$20\%\ of our sample for all VVV bands.

Additionally, we identified a considerable number of variables inside our sample that were missed during the selection of the VIVA catalogue. They are mostly sources with low signal-to-noise ratio, stars with photometric problems, and for about 1\%\ of them, the procedure used to identify the variability failed. These sources can help to improve the variability analysis of the VVV survey and to verify the accuracy of the VVV photometric flags. This information is valuable to improve the procedures and completeness of future releases.

\begin{acknowledgements}
We thank the anonymous referee for the useful suggestions to improve this paper.
We gratefully acknowledge  the use of data from the  ESO Public Survey
program  ID  179.B-2002  taken  with the  VISTA  telescope,  and  data
products from  the Cambridge Astronomical  Survey Unit (CASU).
F. R. H.  thanks  to  Federal  University of  Santa  Catarina  for  the
computational support, and  the IAG/USP and FAPESP program 2018/21661-9 for  the financial support.
R. K. S.   acknowledges   support  from  CNPq/Brazil   through  projects
308968/2016-6 and 421687/2016-9.
D.M.  gratefully acknowledges support provided  by   the  BASAL  Center  for   Astrophysics  and  Associated Technologies (CATA) through grant AFB-170002, and the Ministry for the
Economy,  Development and  Tourism,  Programa Iniciativa  Cient\'ifica
Milenio  grant  IC120009,  awarded  to  the  Millennium  Institute  of
Astrophysics  (MAS),  and from  project  Fondecyt  No. 1170121.
M.C. gratefully acknowledges  additional support by Germany's  DAAD and DFG
agencies, in  addition to FONDECYT grant  \#1171273 and CONICYT/RCUK's
PCI grant DPI20140066.
C.E.F.L. acknowledges a PCI/CNPQ/MCTIC post-doctoral support, 
MCTIC/FINEP (CT-INFRA grant 0112052700), and the Embrace Space Weather Program 
for the computing facilities at INPE.
T.F acknowledges the financial support from the PIBIC CNPq/Brazil.

\end{acknowledgements}

\bibliographystyle{aa}
\bibliography{library}

\begin{appendix}

\section{Merged classes and what they contain}\label{sec:appenA}

In this section we list all VSX\footnote{\url{https://www.aavso.org/vsx/index.php?view=about.vartypes}} types that were merged into the five large classes discussed in Section \ref{sec:varGaia}. The classes include the following types:

\begin{itemize}

\item Eclipsing: E, EA, EB, EP, EW, EC, ED, ESD, AR, D, DM, DS, DW, EL, GS, HW, K, KE, KW, PN, SD, WD.

\item Rotating: ACV, BY, CTTS/ROT, ELL, FKCOM, HB, LERI, NSIN ELL, PSR, R, ROT, RS, SXARI, SXARI/E, TTS/ROT, WTTS/ROT.

\item Pulsating: ACEP, ACYG, AHB1, (B), BCEP, BCEPS, BL, BLAP, BXCIR, CEP, CW, CWA, CWB, CWB(B), CWBS, GWLIB, CW-FO, CW-FU, DCEP, DCEP(B), DCEPS, DCEPS(B), DCEP-FO, DCEP-FU, DSCT, DSCTC, DSCTr, DWLYN, GDOR, HADS, HADS(B), L, LB, LC, LPV, M, O, PPN, PULS, PVTEL, PVTELI, PVTELII, PVTELIII, roAm, roAp, RR, RRAB, RRC, RRD, RV, RVA, RVB, SPB, SR, SRA, SRB, SRC, SRD, SRS, SXPHE, SXPHE(B), V361HYA, V1093HER, ZZ, ZZA, ZZB, ZZ/GWLIB, ZZO, ZZLep.

\item Eruptive: BE, cPNB[e], CTTS, DIP, DPV, DYPer, EXOR, FF, FSCMa, FUOR, GCAS, I, IA, IB, IN, INA, INAT, INB, INS, INSA, INSB, INST, INT, IS, ISA, ISB, RCB, SDOR, TTS, UV, UVN, UXOR, WR, WTTS, YSO, (YY), ZZA/O.

\item Other:  Includes all Cataclismic Variables (AM, CBSS, CBSS/V, CV, DQ, IBWD, N, NA, NB, NC, NL, NL/VY, NR, S, SN, SN I, SN Ia, SN Iax, SN Ib, SN Ic, SN Ic-BL, SN II, SN IIa, SN IIb, SN IId, SN II-L, SN IIn, SN II-P, SN-pec, UG, UGER, UGSS, UGSU, UGWZ, UGZ, UGZ/IW, V, V838MON, VY, ZAND), X-Ray (BHXB, HMXB, IMXB, LMXB, X, XB, XBR, XJ, XN, XP, XPR), and other objects (AGN, APER, BLLAC, CST, GRB, Microlens, MISC, non-cv, NSIN, PER, QSO, S, SIN, Transient, UNC, VAR, VBD).

\end{itemize}

\section{Summary of VSX classes}\label{sec:appenB}

In this section, we list the main features of a few types of variables used in Section \ref{sec:varGaia} along with the full names for all variability classes shown by Fig.~\ref{fig:period_class_MAG_Ks}. The descriptions are taken from the VSX index.

\begin{itemize}

\item BCEP :      $\beta$ Cephei type variables.
\item CWA: W Virginis variables with periods longer than eight days.
\item CWB: W Virginis  variables with  periods shorter than  eight days
  (also known as BL Herculis variables).
\item DCEP:     Classical Cepheids
\item DSCT :     Variables of the $\delta$ Scuti type.
\item HADS:  High-amplitude $\delta$ Scuti stars.
\item M:  o (omicron) Ceti-type (Mira) variables. 
\item MISC: Miscellaneous variables  stars. Usually red variables (L or
  SR)  or  other  types  of  irregular stars,  I or  BE)  that  cannot be
  classified more specifically  by the automatic analysis  made in the
  surveys.
\item PULS:     Pulsating variables of unspecified type. 
\item RR:       Variables of the RR Lyrae type.
\item RRAB: RR Lyrae variables with asymmetric light curves, periods
  from 0.3 to 1.2 days, and amplitudes from 0.5 to 2 mag. in V.
\item  RRC: RR  Lyrae variables  with  nearly symmetric  sometimes
  sinusoidal  light  curves,  periods  from  0.2  to  0.5  days,  and
  amplitudes not greater than 0.8 mag. in V.
\item  RRD:  Double-mode  RR  Lyrae  stars  that  pulsate  in  the
  fundamental mode  as well  as in  the first  overtone with  a period
  ratio of 0.74 and a fundamental period near 0.5 days.
\item RV:       Variables of the RV Tauri type.
\item RVA:       RV Tauri variables that do not vary in mean magnitude.
\item S:  Unstudied variables stars with rapid light changes. 
\item SPB:  Slowly  pulsating B  stars showing  both light  and line
  profile variability.
\item SR: Semi-regular variables, which are giants or supergiants of
  intermediate and late spectral  types showing noticeable periodicity
  in  their light  changes,  accompanied or  sometimes interrupted  by
  various irregularities.
\item SRA:  Semi-regular late-type  (M, C, and S  or Me, Ce,  and Se) giants
  displaying persistent periodicity and usually small (<2.5 mag. in V)
  light amplitudes.
\item SRB: Semi-regular late-type  (M, C, and S  or Me, Ce, and Se) giants
  with poorly defined  periodicity (mean cycles in the range  of 20 to
  2300  days)  or with  alternating  intervals  of periodic  and  slow
  irregular changes, and even with light constancy intervals.
\item  SRS: Semi-regular pulsating  red giants  with short  periods
  (several days to a month), probably high-overtone pulsators.
\item VAR: Variables stars of unspecified type.

\end{itemize}

\end{appendix}

\end{document}